\documentclass[11pt,oneside]{article}
\usepackage{a4wide}
\usepackage{mathrsfs}
\usepackage{latexsym,bm}
\usepackage{graphicx}
\usepackage{subfigure}
\usepackage{indentfirst}
\usepackage{slashed}
\usepackage{amsmath}
\usepackage{amssymb}
\usepackage{color}
\usepackage{hyperref}
\usepackage{epsfig}
\usepackage[titletoc]{appendix}
\usepackage{multirow}%
\usepackage{rotating}
\usepackage{cite}
\usepackage{ulem} 
\usepackage[top=2.9cm,bottom=2.5cm,left=2.8cm,right=3cm]{geometry}
\usepackage{tabularx}
\newcommand{\bma}{\left(\begin{matrix}}
\newcommand{\ema}{\end{matrix}\right)}
\newcommand{\bqa}{\begin{eqnarray}}
\newcommand{\eqa}{\end{eqnarray}}
\newcommand{\bqaa}{\begin{eqnarray*}}
\newcommand{\eqaa}{\end{eqnarray*}}
\newcommand{\mL}{\mathcal{L}}
\newcommand{\mF}{\mathcal{F}}

\newcommand{\mT}{\mathcal{T}}
 
\begin{document}
\thispagestyle{empty}
\title{ \Large \bf Prediction of exotic doubly charmed baryons within chiral effective field theory }
\author{\small Zhi-Hui Guo  \\[0.5em]
{ \small\it  Department of Physics, Hebei Normal University,  Shijiazhuang 050024, China}
}
\date{}

\maketitle

\begin{abstract} 
In this work we study the possible QCD exotic states in the doubly charmed baryon sector. Within chiral effective theory, it is predicted that several excited baryons result from the $S$-wave scattering  of ground-state doubly charmed baryons ($\Xi_{cc}^{++}, \Xi_{cc}^{+}, \Omega_{cc}^{+}$) and light pseudoscalar mesons ($\pi, K, \eta$). The excited doubly charmed baryons can be classified by the strangeness ($S$) and isospin ($I$) quantum numbers. Two of the excited states are clearly exotic, in the sense that they cannot be explained by the conventional baryons with three quarks, since their quantum numbers are $(S,I)=(1,0)$ and $(-1,1)$. Similar to the charmed scalar meson $D_{s0}^{*}(2317)$ in $DK$ scattering and the hyperon $\Lambda(1405)$ in $\bar{K}N$ scattering, one bound state below the $\Xi_{cc} \bar{K}$ threshold is predicted in the $(S,I)=(-1,0)$ channel. In addition, two resonant structures are found in the $\Xi_{cc}\pi, \Xi_{cc}\eta$ and $\Omega_{cc} K$ coupled-channel scattering with $(S,I)=(0,1/2)$. The corresponding pole positions and coupling strengths of the excited doubly charmed baryons are given. The scattering lengths of the ground-state doubly charmed baryons and light pseudoscalar mesons are also predicted. The current study may provide useful guides for future experimental measurements and lattice simulations.

\end{abstract}


\section{Introduction}

Since the discovery of $X(3872)$~\cite{Choi:2003ue}, the study of QCD exotic hadronic states with open/hidden heavy flavors has become the most active research topic in hadron physics. Many comprehensive review articles with different perspectives have appeared to highlight the thriving studies in this research  area~\cite{Chen:2016qju,Chen:2016spr,Esposito:2016noz,Guo:2017jvc,Ali:2017jda}. 
Up to now, the exotic-state studies have covered the hidden heavy flavor mesons and baryons, open heavy flavor mesons and baryons with one heavy quark. Investigations of the exotic baryons with two heavy flavor quarks are still rare. The current work extends the QCD exotic studies in the doubly charmed baryon sector.

The ground-state doubly charmed baryons with spin 1/2 are expected to have three members-- $\Xi_{cc}^{++}$, $\Xi_{cc}^{+}$, and $\Omega_{cc}^{+}$-- with the valence-quark components $ccu, ccd$, and $ccs$, respectively. There is a long-standing puzzle regarding the experimental observation of such states. The SELEX Collaboration has observed the mass of the singly charged state $\Xi_{cc}^{+}$ to be $3519\pm 2$~MeV~\cite{Mattson:2002vu,Ocherashvili:2004hi}. Nevertheless, this state was not seen by the FOCUS~\cite{Ratti:2003ez}, BaBar~\cite{Aubert:2006qw} or Belle~\cite{Chistov:2006zj} collaborations. 
The existence of the doubly charmed baryon $\Xi_{cc}^{++}$ with a mass of $3621.4\pm 0.78$~MeV was recently confirmed by the LHCb Collaboration~\cite{Aaij:2017ueg}. This observation has been the focus of many recent theoretical works~\cite{Chen:2017sbg,Li:2017cfz,Wang:2017azm,Meng:2017udf,Karliner:2017elp,Wang:2017mqp,Kerbikov:2017pau,Gutsche:2017hux,Eichten:2017ffp,Karliner:2017qjm}.

In this work we further investigate the possible exotic states in the scattering of ground-state doubly charmed baryons ($\Xi_{cc}^{++},\Xi_{cc}^{+},\Omega_{cc}^{+}$) and the light pseudoscalar mesons ($\pi, K, \eta$). Chiral effective field theory provides a reliable framework to perform such a study. The leading-order calculation, which includes the Weinberg-Tomozawa/contact interactions, and the $s$- and $u$-channel ground-state exchanges, shall be carried out. The possibly strong interactions between the ground-state doubly charmed baryons and light pseudoscalar mesons are taken into account through the unitarization approach. With this setup, we are able to predict the scattering lengths, phase shifts, line shapes, excited baryon resonance pole positions, and their coupling strengths.

\section{Chiral Lagrangian, partial-wave amplitude, and its unitarization}\label{sec.lagrangian}

The ground-state doubly charmed baryons are expected to form a chiral triplet, 
\begin{eqnarray}\label{eq.xicc}
\psi_{cc} = \left( \begin{array}{c}
 \Xi_{cc}^{++} \\ 
 \Xi_{cc}^{+} \\ 
 \Omega_{cc}^{+} \\ 
\end{array} \right) \,.
\end{eqnarray}
The light pseudoscalar mesons ($\pi, K, \eta$) are identified as the pseudo-Nambu-Goldstone bosons (pNGBs) of QCD, which are introduced into the chiral Lagrangian through~\cite{Gasser:1984gg}
\begin{eqnarray}\label{eq.u}
 U= u^2 &=& e^{i \sqrt2\Phi /F }\,,
\end{eqnarray}
with 
\begin{equation}\label{eq.phi1} 
\Phi \,=\, \left( \begin{array}{ccc}
\frac{1}{\sqrt{2}} \pi^0+\frac{1}{\sqrt{6}}\eta  & \pi^+ & K^+ \\ \pi^- &
\frac{-1}{\sqrt{2}} \pi^0+\frac{1}{\sqrt{6}}\eta & K^0 \\  K^- & \bar{K}^0 &
\frac{-2}{\sqrt{6}}\eta
\end{array} \right)\,.
\end{equation} 
$F$ represents the pion decay constant in the chiral limit. 

The leading-order (LO) chiral Lagrangian consisting of doubly charmed baryons and light pNGBs reads~\cite{Scherer:2002tk,Sun:2014aya,Sun:2016wzh}  
\begin{eqnarray}\label{eq.lolag}
\mL = \bar{\psi}_{cc} \left(i \slashed{D} - m_0 + \frac{g_A}{2} \gamma^\mu\gamma_5 u_\mu \right) \psi_{cc} \,,
\end{eqnarray}
with 
\begin{eqnarray}
D_\mu  =\partial_\mu + \frac{1}{2}\bigg[ u^\dagger \partial_\mu  u
+ u \partial_\mu u^\dagger \bigg] \,,  \,\,
 u_\mu = i u^+ D_\mu U u^+  \,.
\end{eqnarray}
The axial-vector coupling constant $g_A$ was estimated in Ref.~\cite{Sun:2016wzh}. The same value $|g_A|= 0.2$ shall be used here. 
 
\begin{figure}[htbp]
   \centering
   \includegraphics[width=0.95\textwidth,angle=-0]{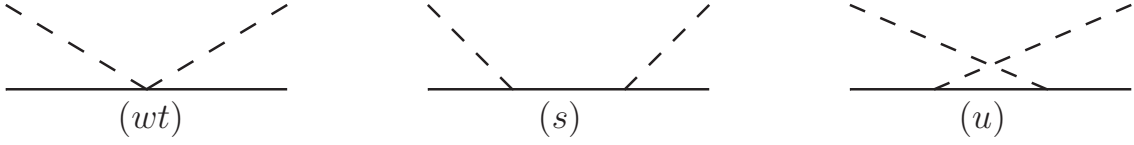} 
  \caption{ Feynman diagrams for the scattering of light pNGBs and doubly charmed baryons. The dashed and solid lines denote the light pNGBs and baryons, respectively. }
   \label{fig.feynmandiagram}
\end{figure}

Three types of Feynman diagrams corresponding to the light pNGBs scattering off the doubly charmed baryons arise from the Lagrangian in Eq.~\eqref{eq.lolag}, which are illustrated in Fig.~\ref{fig.feynmandiagram}.  
The Weinberg-Tomozawa/contact interactions are contributed by the first term in Eq.~\eqref{eq.lolag}, while the $s$- and $u$-channel exchanges are given by the $g_A$ term.

The scattering amplitudes for $\psi_{cc\,,A} (p) \,\phi_i (q) \to \psi_{cc\,,B}(p') \,\phi_j (q')$-- which are the sum of the three diagrams in Fig~\ref{fig.feynmandiagram}-- denoted as $ T^{Ai\to Bj}$, can be calculated using the LO Lagrangian. The explicit expressions are
\begin{eqnarray}\label{eq.t} 
 T^{Ai\to Bj} = T^{Ai\to Bj}_{wt} + T^{Ai\to Bj}_{s} + T^{Ai\to Bj}_{u} \,,
\end{eqnarray}
\begin{eqnarray}\label{eq.tci}
T^{Ai\to Bj}_{wt} &=& - \mF_{wt}\frac{1}{F^2} \,\bar{u}_B (\slashed{q} + \slashed{q}') u_A \,, \nonumber\\
T^{Ai\to Bj}_{s} &=& - \mF_{s}\frac{g^2_A}{F^2}\, \bar{u}_B \slashed{q}' \gamma_5\, \frac{1}{\slashed{p}_C - m_C} \,\slashed{q} \gamma_5 u_A  \,, \nonumber\\
T^{Ai\to Bj}_{u} &=& - \mF_{u}\frac{g^2_{A}}{F^2}\, \bar{u}_B \slashed{q} \gamma_5 \,\frac{1}{\slashed{p}_C - m_C}\, \slashed{q}' \gamma_5 u_A  \,,
\end{eqnarray}
where $p_C$ and $m_C$ stand for the momentum and mass of exchanged baryons in the $s$ and $u$ channels. For the $s$ channel $p_C=p+q$, and for the $u$ channel $p_C= p-q'$. 
The general scattering processes of $\psi_{cc\,,A} \,\phi_i  \to \psi_{cc\,,B}\,\phi_j$ can be decomposed into seven independent channels with definite strangeness and isospin quantum numbers. The symmetry coefficients  $\mF_{wt}$, $\mF_{s}$, and $\mF_{u}$ in Eq.~\eqref{eq.tci} for different processes are summarized in Table~\ref{tab.ci}, where the exchanged particles in the $s$ and $u$ channels are also indicated.

\begin{table*}[htbp]
\centering
{\footnotesize 
\begin{tabular}{l c c c c || l c c c c}
\hline\hline
 $(S,I)$ & Processes & $\mF_{wt}$ & $\mF_{s}$ ($C$) & $\mF_{u}$ ($C$)  &  $(S,I)$ & Processes & $\mF_{wt}$ & $\mF_{s}$ ($C$) & $\mF_{u}$ ($C$)  
 \\
\hline
$(-2,\frac12)$ & $\Omega_{cc}\bar{K}\to \Omega_{cc}\bar{K}$   & $\frac{1}{4}$    & 0 & $\frac{1}{2}$ ($\Xi_{cc}$) & $(1,0)$      & $\Xi_{cc} K\to \Xi_{cc} K$   & -$\frac{1}{4}$    & 0 & -$\frac{1}{2}$  ($\Omega_{cc}$)
\\
$(0,\frac32)$ & $\Xi_{cc}\pi\to \Xi_{cc}\pi$  & $\frac{1}{4}$    & 0 & $\frac{1}{2}$  ($\Xi_{cc}$) & $(1,1)$      & $\Xi_{cc} K\to \Xi_{cc} K$   & $\frac{1}{4}$    & 0 & $\frac{1}{2}$ ($\Omega_{cc}$)
\\
$(-1,1)$       & $\Omega_{cc}\pi\to \Omega_{cc}\pi$    & 0   & 0 & 0  & $(0,\frac12)$ & $\Xi_{cc}\pi\to \Xi_{cc}\pi$  & -$\frac{1}{2}$ & $\frac{3}{4}$ ($\Xi_{cc}$) & -$\frac{1}{4}$  ($\Xi_{cc}$)
\\
              & $\Xi_{cc}\bar{K}\to \Xi_{cc}\bar{K}$   & 0   & 0 & 0 &              & $\Xi_{cc}\eta\to \Xi_{cc}\eta$     & 0   & $\frac{1}{12}$ ($\Xi_{cc}$) & $\frac{1}{12}$ ($\Xi_{cc}$)
\\
              & $\Omega_{cc}\pi\to \Xi_{cc}\bar{K}$    & $\frac{1}{4}$    & 0 & $\frac{1}{2}$  ($\Xi_{cc}$) &              & $\Omega_{cc} K\to \Omega_{cc} K$  & -$\frac{1}{4}$    & $\frac{1}{2}$ ($\Xi_{cc}$) & 0

\\
$(-1,0)$       & $\Xi_{cc}\bar{K}\to \Xi_{cc}\bar{K}$   & -$\frac{1}{2}$    &  $1$ ($\Omega_{cc}$) & 0 &              & $\Xi_{cc}\pi\to \Xi_{cc}\eta$     & 0 & $\frac{1}{4}$ ($\Xi_{cc}$)  & $\frac{1}{4}$ ($\Xi_{cc}$) 

\\
              & $\Omega_{cc}\eta\to \Omega_{cc}\eta$ & 0  & $\frac{1}{3}$  ($\Omega_{cc}$) & $\frac{1}{3}$  ($\Omega_{cc}$) &              & $\Xi_{cc}\pi \to \Omega_{cc} K$ & -$\frac{\sqrt3}{4\sqrt2}$ & $\frac{\sqrt3}{2\sqrt2}$  ($\Xi_{cc}$)  & 0

\\
             & $\Xi_{cc}\bar{K}\to \Omega_{cc}\eta$  & $\frac{\sqrt3}{4}$ & -$\frac{1}{\sqrt3}$  ($\Omega_{cc}$)  & $\frac{1}{2\sqrt3}$ ($\Xi_{cc}$) &             & $ \Xi_{cc}\eta\to \Omega_{cc} K$  & -$\frac{\sqrt3}{4\sqrt2}$   & $\frac{1}{2\sqrt6}$  ($\Xi_{cc}$)  & -$\frac{1}{\sqrt6}$  ($\Omega_{cc}$)             
\\
\hline\hline
\end{tabular} 
\caption{\label{tab.ci} The symmetry coefficients in Eq.~\eqref{eq.tci}. The 
processes are labeled by strangeness ($S$) and isospin ($I$). The particle contents accompanying the coefficients $\mF_{s}$ and $\mF_{u}$ indicate the exchanged doubly charmed baryons. } 
}
\end{table*}

We proceed to perform the partial-wave projections and only the $S$-wave scattering will be considered. The $S$-wave  projection formula for the baryon-meson scattering reads 
\begin{eqnarray} \label{eq.pwt}
 \mT(W)= \frac{1}{8\pi} \sum_{\sigma=1,2} \int d\Omega \, T(W,\Omega;\sigma,\sigma)\,,
\end{eqnarray}
where $W$ is the center-of-mass energy , $\sigma$ denotes the third component of the spin for the incoming and outgoing baryons and $\Omega$ stands for the solid angle of the scattered three-momentum. Notice that the process indices are omitted in Eq.~\eqref{eq.pwt} for simplicity.

Bound states or resonances may appear in the two-body scattering if the interactions are strong enough. A prominent example is $D_{s0}^{*}(2317)$, which can be nicely explained as the bound state from the scattering of charmed mesons and light pNGBs~\cite{Guo:2009ct,Guo:2015dha,Du:2017ttu}. 
Another example is $\Lambda(1405)$. The interactions of the light baryon octet and pNGBs in the strangeness $-1$  channel are so strong that $\Lambda(1405)$ can be naturally generated in their scattering~\cite{Oller:2000fj,Jido:2003cb,Ikeda:2012au,Guo:2012vv}. 
In this work we are interested in exploring whether excited doubly charmed baryon bound states or resonances could appear in the ground-state baryons and light pNGBs scattering. Thus, unitarity plays a key role in this kind of study. A similar unitarization approach following the previous studies on $D_{s0}^{*}(2317)$ and $\Lambda(1405)$ shall be used. This approach is an algebraic approximation of the N/D approach and the final scattering amplitudes respecting  right-hand unitarity read~\cite{Oller:1997ti,Oller:2000fj} 
\begin{eqnarray} \label{eq.defut} 
 \mathbb{T}(s) = \big[ 1 - \mT(s)\cdot G(s) \big]^{-1}\cdot \mT(s)\,,
\end{eqnarray} 
where $s=W^2$ and $\mT(s)$ is the partial-wave amplitude calculated in Eq.~\eqref{eq.pwt}. This formalism has been widely used in many phenomenological  discussions~\cite{Oller:1997ti,Oller:2000fj,Guo:2011pa,Jido:2003cb,Guo:2012yt,Ikeda:2012au,Guo:2009ct,Guo:2012vv,Du:2017ttu,Guo:2015daa,Lu:2014ina,Guo:2015dha}. The function $G(s)$ only contains the right-hand or unitarity cuts contributed by the two-particle intermediate states and is given by the standard two-point one-loop function 
\begin{eqnarray}\label{eq.defg}
G(s)=\frac{1}{i}\int\frac{{\rm d}^4q}{(2\pi)^4}
\frac{1}{(q^2-m_1^2+i\epsilon)[(P-q)^2-m_2^2+i\epsilon ]}\, , \nonumber \\
\end{eqnarray}
where $s=P^2$, and $m_1$ and $m_2$ the masses of the two intermediate states. 
We use the sharp momentum cutoff to regularize the ultraviolet-divergent $G(s)$ function. 
Then it is convenient to integrate out the zeroth component of the four-momentum integration in Eq.~\eqref{eq.defg}, which leads to 
\begin{eqnarray}\label{eq.defg3d}
G^{\rm \Lambda}(s)= -\int^{|\vec{q}|< \Lambda} \frac{{\rm d}^3 \vec{q}}{(2\pi)^3} \, \frac{w_1+w_2}{2w_1 w_2 \,[s-(w_1+w_2)^2]}, 
\end{eqnarray}
where  $w_i=\sqrt{|\vec{q}|^2+m_i^2}$ and $\Lambda$ is the three-momentum cutoff. The explicit expression for $G^{\rm \Lambda}(s)$ was given in Ref.~\cite{Oller:1998hw}.

One can obtain the phase shifts and inelasticities from the $\mathbb{S}$ matrix, which   
is related to the unitarized amplitude $\mathbb{T}$ in Eq.~\eqref{eq.defut} via  $\mathbb{S}  = 1 + 2 i \sqrt{\rho(s)}\cdot \mathbb{T}(s)\cdot \sqrt{\rho(s)}$. 
In the coupled-channel case, the kinematical factor $\rho(s)$ corresponds to the diagonal matrix with the $k$th element $\rho_k(s)=\sqrt{\big[s-(m_{1,k}-m_{2,k})^2\big]\big[s-(m_{1,k}+m_{2,k})^2\big]}/(16\pi s)$, 
where $m_{1,k}$ and $m_{2,k}$ are the masses of the two intermediate states in the $k$th channel. 
The phase shifts $\delta_{kk}$, $\delta_{kl}$, and inelasticities $\varepsilon_{kk}$, $\varepsilon_{k l}$,  for $k\neq l$, are given by $\mathbb{S}_{ k k} = \varepsilon_{k k} {\rm e}^{2 i \delta_{ k k}}$ and 
$\mathbb{S}_{ k l} = i \varepsilon_{k l} {\rm e}^{ i \delta_{ k l}}$.

\section{Results and discussions}\label{sec.result}

To proceed with the phenomenological discussions, the values for the masses of the ground-state doubly charmed baryons and the cutoff scale $\Lambda$ are needed. At LO, it is safe to neglect the isospin-symmetry-breaking effects and one can use the same mass for $\Xi_{cc}^{++}$ and $\Xi_{cc}^{+}$. We take the newly measured value from LHCb. For $\Omega_{cc}^{+}$, we use the value from Ref.~\cite{Chen:2017sbg}, which gives $m_{\Omega_{cc}^{+}}=3700$~MeV. The LO pion decay constant $F$ will be approximated by its physical value $F_\pi=92.1$~MeV~\cite{Olive:2016xmw}.   
For the axial-vector coupling $g_A$, we verify that it plays a minor role in our numerical results; e.g., both $|g_A|=0.2$ and $g_A=0$ are used in our later study and only very slight changes are observed between the two situations. Therefore, we will only show the results with $|g_A|=0.2$.

The three-momentum cutoff $\Lambda$ is the most sensitive parameter in our study. Its precise value can only be  determined if data is available from the scattering of the ground-state doubly charmed baryons and light pNGBs. Unfortunately, neither experimental measurements nor lattice simulations can currently provide relevant data. Common sense tells us that the scale for the three-momentum cutoff $\Lambda$ in hadronic processes is typically around 1~GeV. We take a conservative estimate for $\Lambda$ of $(1.0\pm 0.2)$~GeV. 
Dimensional regularization by replacing the divergence by a subtraction constant can be also used to calculate the $G(s)$ function in Eq.~\eqref{eq.defg}~\cite{Oller:1998hw}. A theoretical way to determine the subtraction constant is to require that the unitarized $u$-channel amplitude has the same input pole as that exchanged in the $s$-channel case~\cite{Kang:2016zmv}. This amounts to imposing the vanishing condition for the corresponding $G(s)$ function at the exchanged pole position. We apply this procedure to the $\Omega_{cc}\eta \to \Omega_{cc}\eta $ process with $(S,I)=(-1,0)$, and $\Xi_{cc}\pi \to \Xi_{cc}\pi $ and $\Xi_{cc}\eta \to \Xi_{cc}\eta $ with $(S,I)=(0,1/2)$, where the exchanged particles in the $s$ and $u$ channels are identical. The average value of the subtraction constants in these three processes is around $-3.42$ at the renormalization scale $\mu=770$~MeV. By requiring the equalities of the $G(s)$ functions evaluated in the cutoff and dimensional regularization schemes at the thresholds, one can obtain the corresponding values for the three-momentum cutoffs in terms of the subtraction constants for different channels. The average cutoff scale $\Lambda$ in the three channels is found to be around $900$~MeV, which is well within our estimated range. This independent approach to determine the cutoff scale through the subtraction constants gives us further confidence about the reasonableness of the estimated region between $0.8$ and $1.2$~GeV.

In our notation, the $S$-wave scattering lengths are related to the unitarized amplitudes $\mathbb{T}(s)$ through 
\begin{eqnarray}
a_{\psi_{cc}\phi \to \psi_{cc}\phi}^{S,I} = \frac{1}{8\pi (m_{\psi_{cc}}+m_\phi)} \mathbb{T}_{\psi_{cc}\phi \to \psi_{cc}\phi}^{S,I}({ s_{\rm thr}})\,, 
\end{eqnarray}
with $s_{\rm thr}= (m_{\psi_{cc}}+m_\phi)^2 $. The scattering lengths from various channels are summarized in Table~\ref{tab.sl}, which could provide useful reference values for future lattice simulations.

 \begin{table*}[htbp]
 \centering
{\footnotesize 
\begin{tabular}{ c c  c || c c c }
\hline\hline
$(S,\, I)$& Processes & Scattering lengths (fm)      & $(S,\, I)$& Processes & Scattering lengths (fm)
\\ \hline
$(-2,1/2)$& $\Omega_{cc}\bar{K}\to \Omega_{cc}\bar{K}$ & $-0.19^{+0.02}_{-0.02}$       & $(-1,0)$ &$\Xi_{cc} \bar{K}\to \Xi_{cc} \bar{K}$ & $-0.49^{+0.10}_{-0.19}$ 
\\
$(1,0)$&$\Xi_{cc} K\to \Xi_{cc} K $ & $5.2, \, -3.6, \,-1.4$          &  &$\Omega_{cc} \eta\to \Omega_{cc}\eta$&$-0.26^{+0.03}_{-0.03}+i\,0.02^{+0.02}_{-0.01}$ 
\\
$(1,1)$&$\Xi_{cc} K\to \Xi_{cc} K $ & $-0.19^{+0.02}_{-0.02}$          & $(0,1/2)$&$\Xi_{cc}\pi\to \Xi_{cc}\pi$ &  $0.55^{+0.16}_{-0.10}$ 

\\
$(0,3/2)$&$ \Xi_{cc}\pi\to \Xi_{cc}\pi$& $-0.095^{+0.003}_{-0.004}$       &  &$ \Xi_{cc}\eta\to \Xi_{cc}\eta$ & $-0.72^{+0.21}_{-0.17}+i\,0.30^{+1.10}_{-0.18}$ 

\\
$(-1,1)$ &$\Omega_{cc}\pi\to \Omega_{cc}\pi $ & $0.03^{+0.01}_{-0.01}$       & &$ \Omega_{cc} K\to \Omega_{cc} K$ &$-0.55^{+0.11}_{-0.16}+i\,0.13^{+0.19}_{-0.07}$ 

\\
 &$\Xi_{cc} \bar{K}\to \Xi_{cc} \bar{K}$&$-0.22^{+0.14}_{-0.14} + i 0.45^{+0.00}_{-0.09}$ 
\\
\hline\hline
\end{tabular}
\caption{\label{tab.sl} Predictions of the scattering lengths using the parameter $\Lambda$ ranging from 0.8 to 1.2~GeV. The central values are obtained with $\Lambda=1.0$~GeV. For the $\Xi_{cc} K\to \Xi_{cc} K $ scattering with $(S,I)=(1,0)$, the scattering length becomes huge when $\Lambda \simeq 0.9$~GeV. As discussed later, the reason for this is that the bound-state pole approaches the $\Xi_{cc} K$ threshold for such a value of $\Lambda$. We simply present the three different scattering lengths obtained at $\Lambda=0.8, 1.0$ and $1.2$~GeV for this channel.  }
}
\end{table*}

We need to perform the analytical continuation of the scattering amplitudes, since the resonance pole is located in the complex energy plane on the unphysical Riemann sheet (RS). This can be realized by the analytical continuation of the $G^{\Lambda}(s)$ function through 
\begin{eqnarray}\label{eq.defg2ndrs}
 G^{\Lambda}_{II}(s) = G^{\Lambda}(s) - 2 i \rho(s)\,,
\end{eqnarray}
where $\rho(s)$ is the kinematical factor introduced previously. $G^{\Lambda}_{II}(s)$ is the expression of the $G^{\Lambda}(s)$ function on the second RS. The physical/first RS is denoted by $(+,+,+,...,+)$, where the plus sign in  each entry corresponds to the sign of the imaginary part of the $G^{\Lambda}(s)$ function at the corresponding threshold. The second RS is labeled by $(-,+,+,...,+)$. The pole positions and their residues obtained in different processes are collected in Table~\ref{tab.pole}.

\begin{table*}[htbp]
 \centering
 {\footnotesize 
\begin{tabular}{ c c | c c c c c}
\hline\hline
$(S,I)$  & RS & Mass(MeV)   & Width/2(MeV)   & $|{\rm Residue}|_{11}^{1/2}$(GeV)  & $|{\rm Residue}|_{22}^{1/2}$(GeV)  & $|{\rm Residue}|_{33}^{1/2}$(GeV)  \\
 \hline
$(1,0)$  & I  & $(--,\,4112,\,4096)$ & $0$ & $(--,\, 10.0 ,\, 14.9)$ &  & \\
  & II & $(4114,\,4115,\,4113)$ & $0$ & $(6.5,\, 3.9 ,\, 4.1)$ &  & \\
\hline
$(-1,1)$ & II & $(4191,\,4134,\,4090)$ & $(89,\,83,\,74)$ & $( 15.7,\, 14.5, \, 13.2)$ &  $( 21.0 ,\, 18.3 ,\,16.4)$ & \\
\hline
$(-1,0)$ & I & $(4018,\,3957,\,3907)$ & $0$ & $(22.4 ,\, 21.7 ,\, 19.8)$ & $( 13.4 ,\, 12.5 ,\, 11.1)$  & \\
        & II & $(4105,\,4095,\,4083)$ & $0$ & $(5.7,\, 6.6 ,\, 7.4)$ & $(3.0,\, 3.2 ,\, 3.3)$  & \\
\hline
$(0,\frac{1}{2})$ & II & $(3830,\,3816,\,3800)$ & $(76,\,50,\,33)$ & $(15.7 ,\, 14.6 ,\,13.4)$ & $(1.0 ,\, 1.2 ,\, 1.2)$ & $(8.3 ,\, 7.6 ,\,6.9)$ \\
 & II & $(4170,\,4146,\,4116)$ & $(8,\,18,\,22)$  & $(4.4 ,\, 5.7 ,\,6.4)$  & $(9.9 ,\, 12.1 ,\, 13.2)$ & $(12.0 ,\, 13.6 ,\,14.1)$ \\
 \hline\hline
\end{tabular}
\caption{\label{tab.pole} Masses and widths of the resonance poles and their residues in the complex energy plane. The three numbers inside each set of parentheses are obtained by taking $\Lambda=(0.8,\, 1.0, \, 1.2)$~GeV. For $\Lambda=0.8$~GeV, we do not find  the bound-state pole on the first RS with $(S,I)=(1,0)$. We verify that around $\Lambda \sim 0.9$~GeV the bound-state pole in this channel approaches rather close to the $\Xi_{cc} K$ threshold.  }
}
\end{table*}

\begin{figure}[htbp]
\centering
\includegraphics[width=0.95\textwidth,angle=-0]{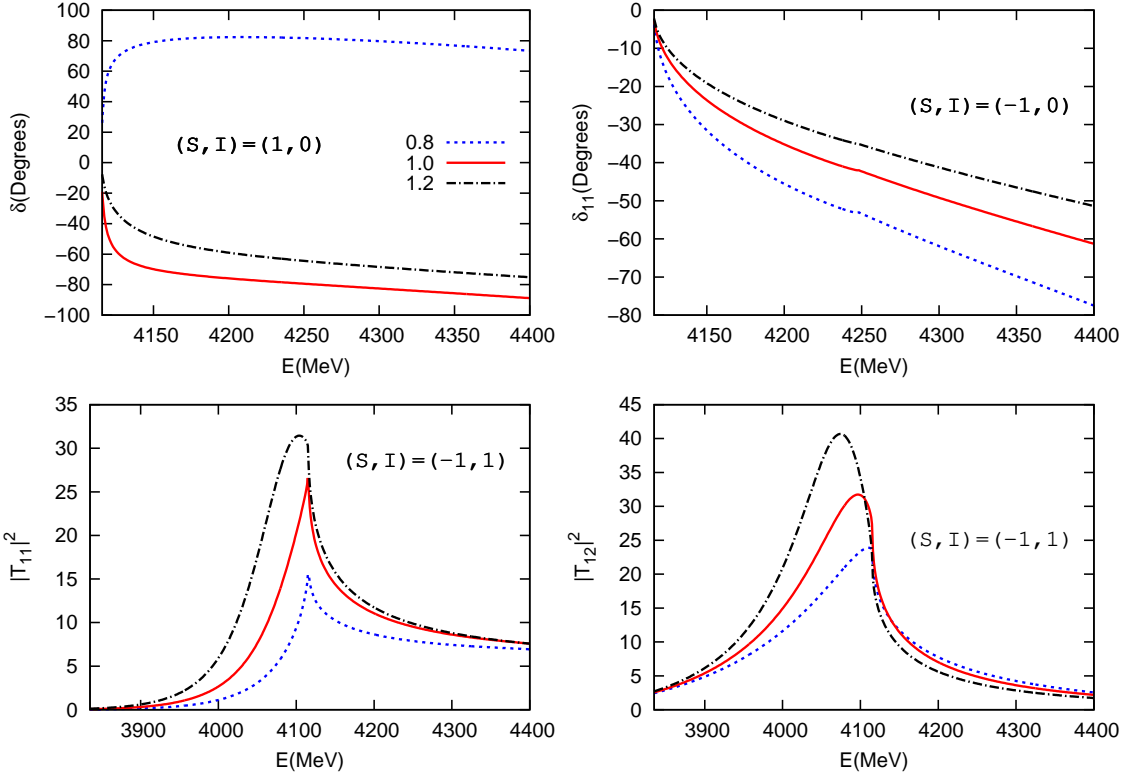} 
\caption{ The upper left and right panels show the phase shifts of $\Xi_{cc}K \to \Xi_{cc}K$ with $(S,I)=(1,0)$ and $\Xi_{cc} \bar{K} \to \Xi_{cc} \bar{K}$ with $(S,I)=(-1,0)$, respectively. The lower two panels show the magnitude squared of the amplitudes from the $\Omega_{cc} \pi \to \Omega_{cc} \pi$ and $\Omega_{cc} \pi \to \Xi_{cc} \bar{K}$ processes. Three different results with the three-momentum cutoff scale $\Lambda=0.8,\, 1.0, \, 1.2$~GeV are shown.  }
   \label{fig.1}
\end{figure}

A pair of bound and virtual states, appear near the threshold in the $S$-wave $\Xi_{cc} K$ scattering with  $(S,I)=(1,0)$, implying a prominent enhancement around the threshold. In Fig.~\ref{fig.1}, the steep increasing/decreasing phase shifts for the $\Xi_{cc} K$ scattering with $(S,I)=(1,0)$ are shown. According to the quantum numbers, the poles clearly cannot be explained by the conventional three-quark doubly charmed baryon.

A resonant pole, with a mass around 4100~MeV and a width around 170~MeV, manifests in the $\Omega_{cc}\pi$ and $\Xi_{cc} \bar{K}$ coupled-channel scattering with $(S,I)=(-1,1)$. The resonant peak can be seen in  Fig.~\ref{fig.1}. According to the quantum numbers, the resonant behavior cannot be due to a doubly charmed baryon with three quarks from the conventional quark model. 

In the $\Xi_{cc} \bar{K}$ and $\Omega_{cc}\eta$ coupled-channel scattering with $(S,I)=(-1,0)$, a bound state around 150~MeV below the $\Xi_{cc} \bar{K}$ threshold has been found. This doubly charmed bound state resembles the $\Lambda(1405)$~\cite{Oller:2000fj,Jido:2003cb,Ikeda:2012au,Guo:2012vv} and $D_{s0}^{*}(2317)$~\cite{Guo:2009ct,Guo:2015dha,Du:2017ttu} poles appearing in the scattering of $\bar{K} N$ and $D K$, respectively. In addition, a virtual-state pole close to the $\Xi_{cc} \bar{K}$ threshold is also found. The phase shifts from $\Xi_{cc} \bar{K} \to \Xi_{cc} \bar{K}$ are shown in Fig.~\ref{fig.1}.

\begin{figure}[htbp]
   \centering
   \includegraphics[width=0.98\textwidth,angle=-0]{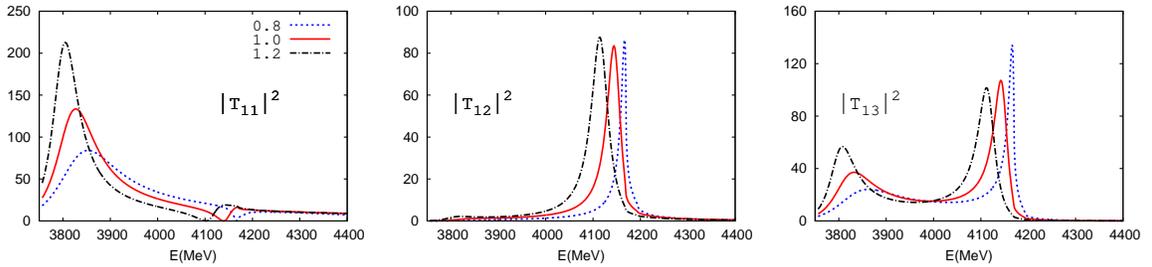} 
  \caption{  From left to right, the three panels give the magnitude squared of the amplitudes for $\Xi_{cc} \pi \to \Xi_{cc} \pi $, $\Xi_{cc} \pi \to \Xi_{cc} \eta $, and $\Xi_{cc} \pi \to \Omega_{cc} K$, respectively. Three different results with the three-momentum cutoff scale $\Lambda=0.8,\, 1.0, \, 1.2$~GeV are shown. } 
   \label{fig.t01d20}
\end{figure}

For the $\Xi_{cc} \pi$, $\Xi_{cc} \eta$ and $\Omega_{cc} K$ coupled-channel scattering, we see two clear resonant poles. The lower pole lies slightly above the $\Xi_{cc} \pi$ threshold, and its mass and width are around 3800 and 100~MeV, respectively. The higher pole lies around the two close thresholds of $\Xi_{cc} \eta$ and $\Omega_{cc} K$, and  its mass and width are around 4150 and 35~MeV, respectively. According to their coupling strengths in Table~\ref{tab.pole}, the lower pole is very weakly coupled to the $\Xi_{cc}\eta$ channel. For the higher pole, its coupling strengths to the $\Xi_{cc} \eta$ and $\Omega_{cc} K$ channels are similar, with the coupling to $\Xi_{cc} \pi$ being only around 1/2 of the former two channels. The line shapes of the scattering processes of $\Xi_{cc} \pi \to \Xi_{cc} \pi$, $\Xi_{cc} \pi \to \Xi_{cc} \eta$ and $\Xi_{cc} \pi \to \Omega_{cc} K$ , are given in Fig.~\ref{fig.t01d20}. Interestingly, we find that the higher  resonance pole around 4150~MeV shows up as a dip, instead of a peak, in the $\Xi_{cc} \pi \to \Xi_{cc} \pi$ channel. In contrast, a prominent peak around 3800~MeV appears in the $\Xi_{cc} \pi \to \Xi_{cc} \pi$ scattering. For the $\Xi_{cc} \pi \to \Xi_{cc} \eta$ process, only the resonant peak around 4150~MeV manifests and the resonance pole around 3800~MeV barely shows up in this channel. 
For the $\Xi_{cc} \pi \to \Omega_{cc} K$ scattering, both resonant peaks clearly show up in the regions of 3800 and 4150~MeV.

\section{Conclusions}\label{sec.conclusion}

In this work, the leading-order chiral scattering amplitudes of the ground-state doubly charmed baryons and the light pseudo-Nambu-Goldstone bosons were calculated and then unitarized within the algebraic approximation of the N/D approach. To take the mass of $\Xi_{cc}$ from the new LHCb measurement and to estimate the mass of $\Omega_{cc}$ and the axial-vector coupling from the literature, we carried out the comprehensive phenomenological discussions by estimating the three-momentum cutoff scale introduced in the unitarization procedure around $0.8 \sim 1.2$~GeV.

With these setups, a pair of bound and virtual poles appear around the $\Xi_{cc} K$ threshold with $(S,I)=(1,0)$ quantum numbers. One resonance pole with a mass and width around 4100 and 170~MeV, respectively shows up in the coupled-channel scattering of $\Omega_{cc} \pi$ and $\Xi_{cc} \bar{K}$. According to the quantum numbers, the former two kinds of resonant enhancements cannot be due to the conventional doubly charmed baryons composed of three quarks. Therefore, the discovery of these two kinds of states would be a clear evidence for the existence of the exotic doubly charmed baryons.

Two prominent resonance peaks were found around the $\Xi_{cc} \pi$ and $\Omega_{cc} K$ (very close to $\Xi_{cc} \eta$) thresholds in the $(S,I)=(0,1/2)$ channel. Very interestingly, one bound state of doubly charmed baryon appears below the $\Xi_{cc} \bar{K}$ threshold with $(S,I)=(-1,0)$ quantum numbers, the dynamics of which look quite similar to  $\Lambda(1405)$ in the $\bar{K} N$ scattering and $D_{s0}^{*}(2317)$ in the $DK$ scattering.

\section*{Acknowledgements}
I would like to thank Jos\'e Antonio Oller for reading the manuscript. 
This work is supported in part by the NSFC under Grants 
No.~11575052, the Natural Science Foundation of Hebei Province under Contract No.~A2015205205,  
grants from the Education Department of Hebei Province under contract No.~YQ2014034, 
and grants from the Department of Human Resources and Social Security of Hebei Province with contract No.~C201400323.

\end{document}